\documentclass[sigconf, 10pt,  nonacm]{acmart}
\usepackage{graphicx} 
\usepackage{array}      
\usepackage{xcolor}

\title{Rethinking Agent Security as a Networking Problem}
\author{Van Tran}
\authornote{These authors contributed equally to this work.}
\orcid{0000-0002-0835-8598}
\affiliation{
    \institution{University of Chicago}
    \city{Chicago}
    \country{United States}
}

\author{Taveesh Sharma}
\authornotemark[1]
\orcid{0000-0002-2616-0390}
\affiliation{
    \institution{University of Chicago}
    \city{Chicago}
    \country{United States}
}

\author{Tajveer Singh Dhesi}
\authornotemark[1]
\orcid{0009-0009-7384-1170}
\affiliation{
    \institution{University of Chicago}
    \city{Chicago}
    \country{United States}
}
\author{Nick Feamster}
\orcid{0000-0001-9315-5201}
\affiliation{
    \institution{University of Chicago}
    \city{Chicago}
    \country{United States}
}

\begin{document}
\begin{abstract}
  AI agents are rapidly becoming more capable and widely deployed, promising substantial gains in productivity and enabling new classes of applications. However, their growing autonomy also introduces significant privacy and security risks. Existing defenses are predominantly agent-centric, relying on the agent itself to detect threats and enforce privacy and security policies. This approach is fundamentally limited because it entrusts policy enforcement to AI agents whose LLM-driven behavior is inherently nondeterministic and vulnerable to manipulation through attacks such as prompt injection. As a result, current defenses cannot reliably prevent privacy and security threats, highlighting a critical need for a new solution to securing AI agent systems.

  The networking community has long grappled with similar challenges and offers insightful principles we can borrow to design a more secure AI agent system. These include centralized control with distributed enforcement, capability-based access for mediating requests to sensitive resources, and least privilege through zero-trust enforcement. Historically, these principles have provided strong deterministic guarantees for networked systems. However, these principles alone are insufficient for AI agents because the safety and appropriateness of an agent's actions often depend on semantic context beyond the expressiveness of static rules.

  Building on these principles, we advocate for a systematic approach to AI agent security that combines deterministic enforcement mechanisms, which provide strong security guarantees, with semantic, context-aware policies that enable nuanced decision-making. We then present a reference architecture and identify key research questions and future directions to guide the design of secure and privacy-preserving AI agent systems.

\end{abstract}

\maketitle

\section{Introduction}
AI agents are LLM-based systems that can autonomously plan, reason, and execute multi-step workflows \cite{yao2023react}. These agents can call APIs, invoke
external tools, access enterprise resources, coordinate with other
agents, and act with limited human oversight, making them a new layer of
networked systems that mediate interactions among users, services, and external APIs \cite{mcp2024,a2a2025}. A
supervisor agent handling a scheduling request, for example, may share calendar details with a third party booking service,
all without human review.

This autonomy introduces potential privacy and security risks. Agents can
inadvertently expose sensitive data such as passwords and personal information through their outputs \cite{elyagoubi2026agentleak, zharmagambetov2025agentdam}, tool calls \cite{elyagoubi2026agentleak, alizadeh2025simple}, or
communication with other agents \cite{elyagoubi2026agentleak, naik2026omni, louck2025improving}. Additionally, adversarial techniques such as (indirect) prompt
injection \cite{greshake2023injection,debenedetti2024agentdojo, zhan2024injecagent, evtimov2026wasp}, memory poisoning \cite{chen2024agentpoison, dong2026memory, yazdinejad2026temporal}, and goal hijacking \cite{jia2024taskshield, deng2025ai, zhang2024towards, jha2025breaking} can further manipulate an
agent's reasoning, causing it to bypass security policies and perform harmful actions such as launching attacks \cite{tan2026prompt, zhu2025cve}, escalating privileges \cite{csa2026overprivileged, thehackernews2026authbypass}, tampering with databases \cite{freestone2026aiagents}, or performing unauthorized financial transactions \cite{patlan2025real}.

Unfortunately, existing security architectures are built around static,
predefined rules \cite{sandhu1996rbac,hardt2012oauth} and are
not designed to address the dynamic, context-dependent behavior of AI agents
\cite{oswal2026nondeterministic, yao2023react}. For example, traditional access control mechanisms authenticate a principal once and authorize subsequent actions based on predefined permissions, without considering the context in which an agent performs those actions \cite{sandhu1996rbac,hardt2012oauth}.
In contrast, agent-level defenses such as using structured querying \cite{chen2025struq}, fine-tuning \cite{chen2025secalign, wallace2024instruction, lan2025cirl}, and privacy-aware prompting \cite{zharmagambetov2025agentdam,lan2025cirl}, attempt to reason about context, intent, and privacy risks. 
Because this enforcement depends on LLM reasoning, it inherits the limitations of LLMs: their behavior is inherently nondeterministic, susceptible to attacks such as prompt injection \cite{mccauley2025samemodel,yao2023react,greshake2023injection}, and can be bypassed by adversarial inputs \cite{jia2025critical,wang2026sok,hackett2025bypassing}. As a result, they cannot provide strong security guarantees.


\noindent\textbf{Proposal:} In this paper, we argue that the network is the natural place to enforce privacy and security controls for AI agents. We propose a network-based security architecture built on two complementary abstractions: deterministic enforcement, which constrains the actions an agent is permitted to perform regardless of whether its reasoning has been compromised, and context-aware filtering, which determines whether an action or disclosure is appropriate given its broader context.
 We instantiate these as a reference architecture
that pairs each agent with a sidecar mediating every request, inter-agent message and
external call as allow, redact, or deny, governed by a control plane outside
the agent's reach.



\section{Existing Defenses \& Limitations}
In this section, we examine traditional security architectures and existing AI agent privacy and security defenses, highlighting why they are insufficient for securing autonomous AI agents. We then identify the architectural capabilities needed to address these gaps.

\subsection{Traditional Security Architectures} 
Traditional cybersecurity defenses are built on the assumption that system behavior can be modeled, constrained, and verified through predefined policies \cite{oswal2026nondeterministic,saltzer1975protection}. Accordingly, existing security architectures rely on deterministic entities, static permissions, and predictable interactions, where deviations from expected behavior are treated as potential security violations \cite{denning1987intrusion, oswal2026nondeterministic}. However, these assumptions break down for AI agents, whose actions are generated through dynamic reasoning processes and shaped by interactions with external tools, APIs, and other agents \cite{lu2026clawless, narajala2025securing}. As a result, agent behavior is inherently context-dependent and difficult to anticipate \cite{yao2023react, narajala2025securing}, making security architectures based on static policies and fixed trust boundaries insufficient for protecting AI agents \cite{rose2020zerotrust, oswal2026nondeterministic}.


Access control provides a concrete example of this architectural mismatch. Traditional access control mechanisms, including role-based access control (RBAC) \cite{sandhu1996rbac} and token-based authorization such as API keys \cite{hardt2012oauth}, are fundamentally identity-centric. A principal is authenticated, granted a fixed set of permissions, and subsequent actions performed under that identity are authorized based on these predefined privileges. While effective for conventional applications, these mechanisms provide limited visibility into the context surrounding an action: the sensitivity of the data being accessed, the purpose of the request, the history of prior interactions, or the downstream recipients of the information.
This limitation becomes particularly problematic for AI agents. An agent may legitimately have access to sensitive resources, yet still violate privacy or security expectations by retrieving unnecessary information, sharing data with inappropriate external tools, or taking actions inconsistent with user intent \cite{shao2024privacylens,zharmagambetov2025agentdam,elyagoubi2026agentleak}.

\subsection{AI Agent Privacy and Security Defenses}

The attack surface of AI agents is significantly broader than that of traditional software systems. AI agents continuously interact with external tools, memory systems, internal resources, and other agents, incorporating information from these sources into their reasoning process. Each interaction point can therefore introduce new opportunities for attacks. For example, prompt injection \cite{greshake2023injection,debenedetti2024agentdojo,zhan2024injecagent,evtimov2026wasp} embeds malicious instructions in retrieved documents or tool outputs to manipulate an agent's behavior; memory poisoning \cite{chen2024agentpoison,dong2026memory,yazdinejad2026temporal} corrupts an agent's persistent memory with false information or malicious instructions; and goal hijacking \cite{jia2024taskshield,deng2025ai,zhang2024towards,jha2025breaking} diverts an agent from the user's intended objective by altering its planning process. Beyond adversarial manipulation, AI agents can introduce privacy and security risks even when operating as intended. Agents may disclose sensitive information in inappropriate contexts \cite{elyagoubi2026agentleak, 10.1145/3786582.3786839, luo2026agentauditor} or perform actions that exceed the intended scope of a user's request \cite{qu2026overeager,  jones2026benign}, creating risks of unintended data exposure or unauthorized operations.



Prior work defends against privacy and security threats in AI agents through a variety of broad strategies.
Reasoning-based defenses \cite{zharmagambetov2025agentdam, lan2025cirl, debenedetti2024agentdojo} aim to guide agents toward privacy- and security-aware decisions by embedding policies into the agent's reasoning process. This can be achieved either through prompting \cite{zharmagambetov2025agentdam}, which provides security instructions, policy guidance, or reasoning steps at inference time, or through model fine-tuning \cite{lan2025cirl}, which aligns the model with privacy and security norms during training. For example, AgentDAM \cite{zharmagambetov2025agentdam} uses chain-of-thought prompting to encourage reasoning about data minimization before taking actions, while CI-RL \cite{lan2025cirl} uses reinforcement learning to align agents with contextual integrity norms.



Filtering-based defenses \cite{zhou2025rescriber, rebedea2023nemo, lakera2023guard, bagdasarian2024airgapagent, jia2024taskshield} inspect information flows to prevent sensitive data or malicious instructions from propagating through an agent system. Depending on the threat model, filtering can be applied at different stages, including agent inputs and outputs \cite{rebedea2023nemo} and communication channels between agents or components \cite{bagdasarian2024airgapagent, jia2024taskshield}. These defenses typically rely on rules or classifiers \cite{rebedea2023nemo, lakera2023guard} to detect, block, redact, sanitize, or validate information before it is processed, shared, or stored, thereby reducing the risk of prompt injection, data leakage, and memory poisoning.

Other defenses, such as information flow and tool control \cite{shi2025progent, costa2025securing, zhong2025rtbas}, sandboxing \cite{wu2024isolategpt, debenedetti2025defeating, meng2025cellmate, piao2025agentbay}, and action verification \cite{miculicich2025veriguard, xiang2024guardagent}, aim to constrain agent behavior through predefined interfaces, restricted execution environments, and approval mechanisms. For example, sandboxing \cite{wu2024isolategpt, debenedetti2025defeating, meng2025cellmate, piao2025agentbay} isolates agents from sensitive resources and external systems, limiting their ability to perform harmful actions or access protected assets.

While agent-level defenses can mitigate specific threats, they are fundamentally limited by their reliance on the agent itself for security decisions. Because LLM-based reasoning is inherently nondeterministic, approaches that depend on agents to interpret and enforce privacy and security policies cannot provide strong enforcement guarantees. Moreover, agents remain vulnerable to attacks such as prompt injection, which can manipulate their behavior and bypass these defenses. Therefore, securing AI agents requires mechanisms that (1) provide deterministic enforcement for critical, high-stakes actions and (2) operate independently from the agent executing those actions.

\textbf{Why Borrow from Networking?} The networking community has spent decades addressing the problem of securing communication among autonomous, diverse, and untrusted endpoints \cite{sherry2012aplomb, sherry2015blindbox, sherry2015ftmb, ballani2005offbydefault, yang2005tva, yaar2004siff, casado2007ethane}. Instead of assuming that every endpoint is trustworthy, networks enforce trust at the boundaries between components using principles such as least privilege, mediated communication, and continuous traffic monitoring. These capabilities are largely absent from endpoint focused agent defenses, which lose visibility once data leaves an individual agent and is sent to other agents, tools, or services. Moreover, endpoint-focused defenses such as prompt-level guardrails, model finetuning and model-side filters have repeatedly been shown to be evadable \cite{jia2025critical,wang2026sok,hackett2025bypassing} in prior work. Because multi-agent AI systems are fundamentally distributed systems of communicating entities, networking provides a natural foundation for securing their interactions.

\section{Design Principles for Secure AI Agents}\label{sec:principles}

\begin{figure}[t]
    \centering
    \includegraphics[width=\linewidth]{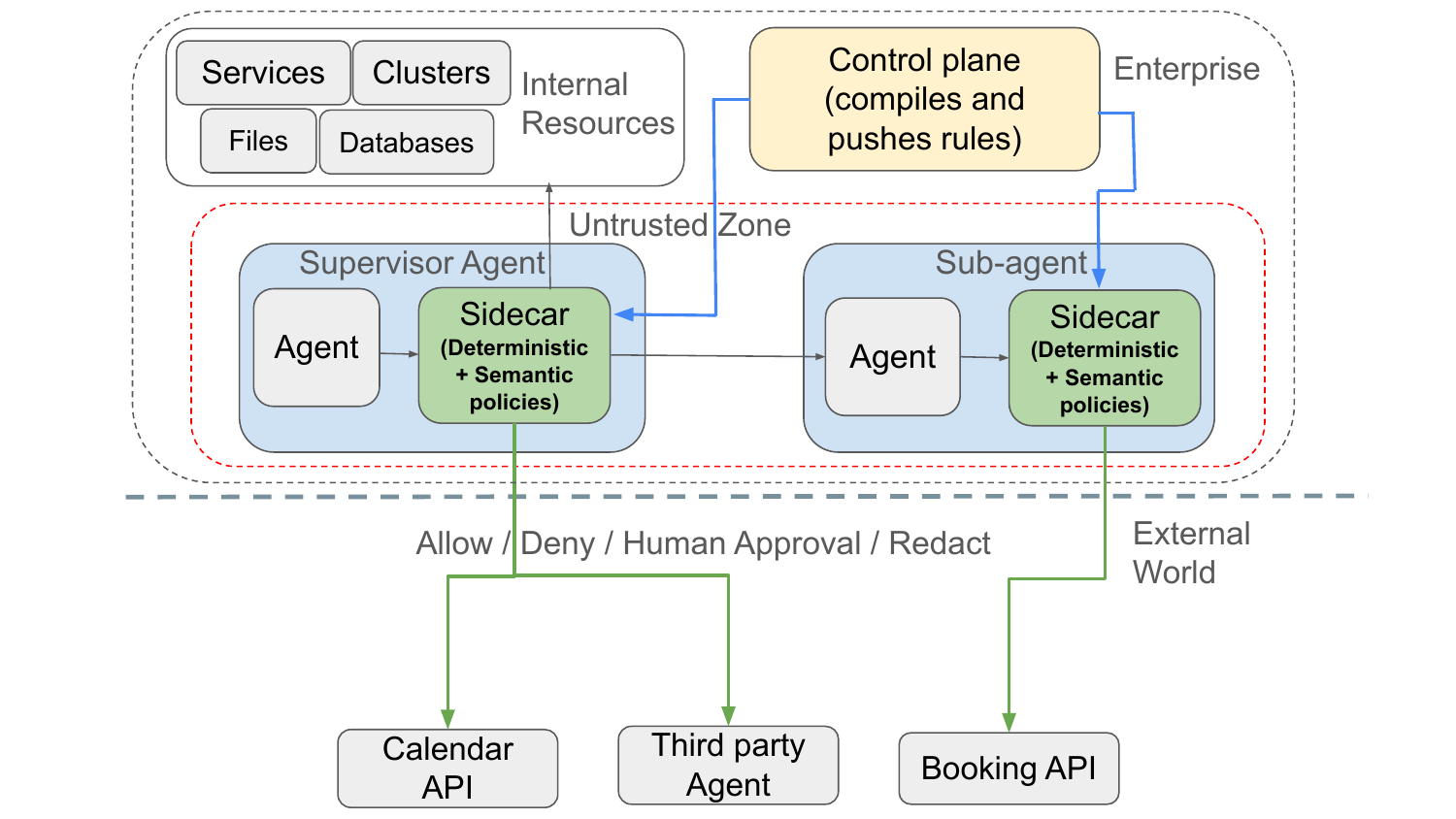}
    \caption{Proposed sidecar architecture for securing multi-agent systems. Each agent is paired with a sidecar that enforces deterministic and semantic policies on all inter-agent messages and external API calls, mediating every action as \emph{allow}, \emph{redact}, \emph{deny} or escalate for the user's \textit{approval}. Agents reside in an untrusted zone, while a control plane in a trusted zone within the enterprise compiles and pushes policy rules to the sidecars, keeping policy authority out of reach of the untrusted agents.}
    \Description{Diagram showing each agent paired with a sidecar enforcing policies on messages and API calls, with a control plane managing policy rules.}
    \label{fig:architecture}
\end{figure}

AI agents are becoming an important component of networked systems, yet they face privacy and security risks that existing network security architectures were never designed to address. Today's defenses for privacy in multi-agent LLM systems reside almost entirely at the application and endpoint layer, in the form of prompt-level guardrails \cite{rebedea2023nemo}, output filters \cite{markov2023holistic}, system-prompt instructions \cite{wallace2024instruction}, and per-agent content classifiers \cite{inan2023llamaguard}. This work argues that the natural place to enforce data minimization and egress control is instead the network, and that the networking community is well positioned to conduct research on agent-aware middleboxes, policy-driven egress filtering, and capability-based network access for agents.

Each of these ideas has a long and well-tested history for hosts, flows, and services. The novelty lies in recognizing that an agent is a new kind of network entity, that inter-agent and agent-to-service traffic is a new kind of flow, and that the privacy failures specific to multi-agent systems (such as benign oversharing, context propagation across agent hops, and contextual integrity violations) are a new kind of failure source. The remainder of this section identifies the components they lack for AI agents, and proposes a set of design principles for building effective agent security architectures. We show that the field has already built the necessary mechanisms, only for a different entity.

\subsection{Combine Deterministic Enforcement with Context-Aware Control}
\label{sec:abstractions}

Our core claim is that agent security requires two abstractions that are useful only in combination: \emph{deterministic execution enforcement} and \emph{context-aware policy semantics}. Deterministic enforcement without context-aware semantics reduces to coarse connectivity control, i.e, it can stop an agent from reaching an unapproved endpoint, but not from oversharing with an approved one. Context-aware semantics without deterministic enforcement reduces to present-day guardrails: norms specified in a prompt and enforced by the very entity being protected. Our contribution is to combine them: compiling semantics into a deterministic mechanism placed where the protected endpoint cannot bypass it.

The enforcement abstraction is a systems-layer primitive that constrains what an agent execution may do. It includes: which tools it may call, which destinations it may reach, which data-flow paths its outputs may traverse. These are agnostic to the agent's cooperation, prompt, or model weights. It resembles a firewall whose unit of protection is the agent execution graph rather than the packet 5-tuple, and it must sit at a network choke point off-path from the agent's reasoning. We develop the mechanism that supplies these constraints in \S\ref{sec:related-capabilities}. The semantic abstraction supplies the vocabulary of appropriateness that principals, resources, and ports cannot. Nissenbaum's \emph{contextual integrity} (CI) provides one such vocabulary~\cite{nissenbaum2004ci}: flows are governed by norms of the form (sender role, recipient role, subject, information type, transmission principle), and a flow violates privacy when it breaches its context's norms. This is why medical context appropriate for a health application becomes a violation when it reaches a restaurant-booking service. CI has been operationalized primarily as an \emph{evaluation} tool in existing defenses, with benchmarks showing high leakage rates~\cite{mireshghallah2024confaide,shao2024privacylens,bagdasarian2024airgapagent}. Recent defenses do apply CI at runtime~\cite{wang2025privacyinaction, lan2025cirl} but they delegate norm application to a model, leaving enforcement probabilistic. We take a different approach: using CI for \emph{enforcement} by turning its norms into checks a choke point can execute. For example, a \texttt{check\_flow(sender, recipient, task, labels)} function evaluated on every hop which, in essence, is information flow control~\cite{myers1997dlm,zeldovich2006histar,krohn2007flume} with CI norms as the label policy and the network as the reference monitor.

\subsection{Centralized Control}
\label{sec:related-centralized}

Connectivity should be governed by a declarative, centrally specified policy that the network enforces, rather than by endpoint mechanisms. This is the thesis of SANE~\cite{casado2006sane} and Ethane~\cite{casado2007ethane}. SANE replaces a mix of access control lists (ACLs), firewalls, and NATs with a single protection layer in which a logically centralized server makes all access-control decisions and implements declarative policies such as ``host $A$ can reach service $B$''; switches are minimally trusted and simply enforce the capabilities they are given. Ethane, which laid the foundation for software-defined networking (SDN), extends this to network-wide policy enforcement using a centralized controller that manages simple flow-based switches. Its central argument is that network policies are effective only if all traffic passes through the enforcement point. The same principle applies to AI agents: if an agent can bypass or ignore a guardrail, the guardrail cannot reliably enforce policy. We adopt Ethane's approach of compiling high-level declarative policies into enforceable rules, extending it from networks to agent egress control.

These parallels raise a question the end-to-end argument was written to answer~\cite{saltzer1984e2e}: when is it appropriate to place a function in the network rather than at the endpoints? The argument is often summarized as ``keep the network dumb,'' but its actual claim is more nuanced: place a function at the endpoints when the endpoints are the only place it can be implemented correctly and completely. Our proposal follows this principle rather than challenging it. The test the argument prescribes is whether the endpoint can do the job completely, and for agent egress control the relevant endpoint is the agent itself. An untrusted and manipulable agent cannot reliably block its own leaks, because the enforcer and the constrained party are the same. Centralized control instead sits outside the agent's boundary, observes every outbound flow, and leaves no part of the job to be completed elsewhere.

\subsection{Capability-based Access}
\label{sec:related-capabilities}

An agent's reach should be off by default: it must obtain an explicit capability granted by policy before it can send data or contact a destination. This idea builds on capability-based networking, developed to control unwanted network traffic. SIFF~\cite{yaar2004siff} separates traffic into privileged and unprivileged classes and requires a capability exchange before privileged communication can begin, allowing receivers to block unwanted traffic before it consumes network resources. The traffic validation architecture (TVA)~\cite{yang2005tva} extends this by requiring senders to first obtain permission from the receiver in the form of capabilities, which network devices verify before discarding unauthorized traffic. The Off-by-Default~\cite{ballani2005offbydefault} proposal applies the same principle: communication is denied by default and allowed only through explicit policy.

We carry this idea over to AI agents, where each task grants a narrow, revocable set of capabilities. By default, an agent should not be able to send a user's data to another agent, tool, or external service. Instead, it should receive a capability that explicitly specifies the allowed destination, the type of data that may be shared, and the purpose of the transfer. As in the networking proposals, these capabilities should be forgery-proof, verifiable by enforcement points, and issued by a trusted policy authority rather than claimed by the agent itself. An open research question is what the right granularity and semantics of such capabilities should be when the protected resource is contextually private information rather than network bandwidth.

\section{Designing the Building Blocks for Secure AI Agents}

Building on the design principles presented in Section \ref{sec:principles}, we present a reference security architecture that combines deterministic enforcement with context-aware semantic control. We first describe the key architectural components and how they interact to enforce security and privacy policies during agent execution. We then discuss how such an architecture can be integrated into existing enterprise environments. We emphasize that this architecture is intended as a reference design rather than a definitive solution. Its purpose is to illustrate the architectural capabilities needed to secure autonomous AI agents, motivate further research, and provide a foundation for developing practical, deployable systems.

\subsection{Core Architectural Components}
At a high level, the architecture separates security policy management from runtime enforcement through two components: a control plane and sidecars. The control plane defines and distributes security policies, while sidecars, deployed alongside AI agents, enforce these policies by mediating agent interactions with external resources. This separation provides centralized policy control while allowing security mechanisms to evolve independently of agent implementations.
\subsubsection{Centralized Control Plane}
The control plane specifies and manages security policies but does not participate in runtime decision making. Located within the enterprise trust boundary and isolated from AI agents, it prevents agents from modifying or bypassing security requirements. It distributes policies to sidecars, including deterministic rules (e.g., ACLs, and RBAC policies), semantic policies (e.g., constraints on how customer PII may be used), request routing policies, data sensitivity labels, risk thresholds, and approved models for semantic evaluation.

\subsubsection{Sidecars} 

Sidecars serve as runtime security enforcement points between AI agents and external resources. They intercept agent actions, evaluate requests using policies received from the control plane, and enforce the resulting decisions before allowing execution. Each sidecar consists of several components responsible for request classification, deterministic enforcement, and context-aware semantic evaluation.

\textbf{Agent Policy Gateway.}
The agent policy gateway serves as the runtime enforcement point between AI agents and external resources. Rather than allowing agents to directly invoke tools or access data, all external actions, including tool invocations, memory operations, API requests, network communications, file access, and data sharing, are intercepted by the gateway before execution. This centralized enforcement point ensures that every action is evaluated against enterprise security policies before it is executed.

\textbf{Request Classifier/ Policy Router.} A key challenge in securing AI agents is that not all actions require semantic reasoning. The request classifier addresses this challenge by determining the appropriate evaluation path for each agent action. Using policy-defined criteria such as action type, resource sensitivity, and risk level, it routes requests either to the deterministic enforcement engine or to the context-aware semantic engine. This risk-adaptive routing enables the architecture to preserve the efficiency and strong guarantees of deterministic enforcement while reserving more expensive semantic analysis for actions whose security implications depend on context, intent, or purpose.

\textbf{Deterministic Enforcement.} The deterministic enforcement engine evaluates requests against predefined, unambiguous security policies that can be enforced without semantic reasoning. These policies include mechanisms such as access control (e.g., ACLs and RBAC), tool permissions, network policies, and rate limits. By relying on explicit rules, the engine enables fast, predictable enforcement without the overhead of semantic reasoning.

\textbf{Context-aware control.} The context-aware semantic engine evaluates requests whose security depends on the surrounding context rather than predefined rules alone. Before making a decision, it gathers the information needed to understand the request, including the sensitivity of the data, the risk of the requested action, the agent's role and permissions, the user's intent, the current task, and the agent's recent interactions with tools and resources. Using this structured view of the agent's execution, the engine evaluates semantic policies to determine whether the request is consistent with its intended purpose and enterprise security requirements. It can then allow, deny, modify, or escalate the request for human approval.

\begin{figure}[t!]
    \centering
    \includegraphics[width=\linewidth]{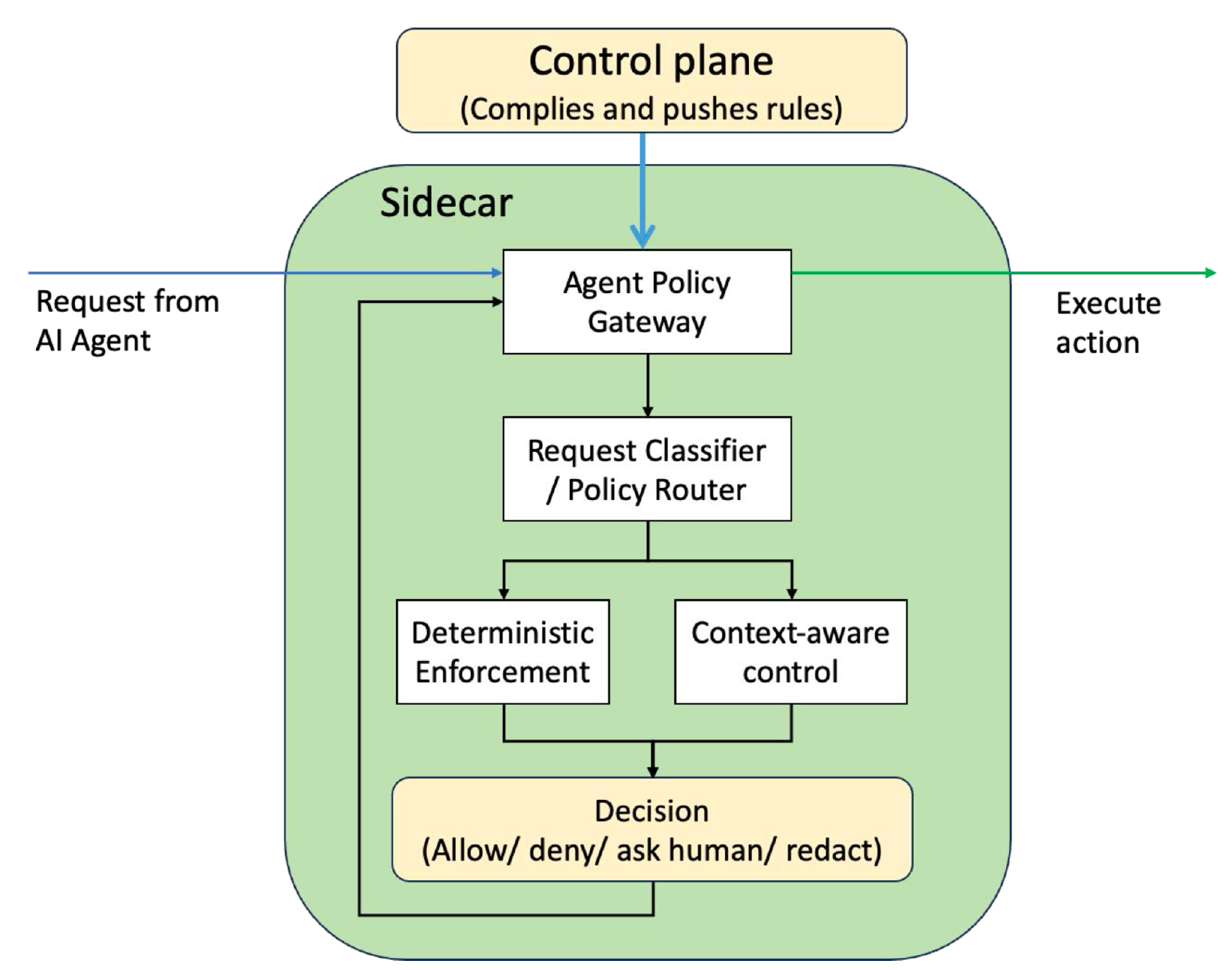}
    \caption{Reference architecture for secure AI agents. The control plane centrally manages security policies, while sidecars deployed at agent interaction points intercept agent actions, route requests to deterministic or semantic enforcement, and mediate access to enterprise resources.}
    \Description{execution flows}
    \label{fig:execution-flow}
\end{figure}

\subsection{End-to-End Execution Flow}

Figure \ref{fig:execution-flow} illustrates the end-to-end execution flow of the proposed architecture. The enterprise control plane first defines security policies and distributes them to the runtime enforcement layer. These policies specify deterministic rules (e.g., access control, and tool permissions), semantic policies for context-dependent decisions, routing policies, and relevant metadata such as data sensitivity labels and risk thresholds.

When an AI agent attempts to perform an external action, such as invoking a tool, accessing memory, issuing an API request, or sharing data, the request is intercepted by the Agent Policy Gateway. The gateway evaluates the request using the policies received from the control plane and forwards it to the Request Classifier, which determines the appropriate enforcement path. Requests with clear, rule-based decisions are routed to the Deterministic Enforcement Engine, while requests requiring reasoning about intent or execution context are routed to the Context-Aware Semantic Engine.

For semantic evaluation, the engine collects relevant execution context, including the requested data, action risk, agent permissions, user intent, task state, and prior interactions. It then evaluates the semantic policies defined by the control plane to determine the appropriate decision. The resulting decision—allow, deny, modify, or escalate for human approval—is returned to the Agent Policy Gateway, which enforces the decision before the action is executed. All decisions and relevant execution information are recorded for auditing and policy refinement.

\subsection{Integration with Existing Architectures}

The proposed architecture enables existing policy enforcement mechanisms to be extended to autonomous AI agents. Similar to API gateways, service meshes, and policy enforcement points, the sidecar mediates agent interactions with protected resources by intercepting and evaluating actions before they are executed. The control plane integrates with existing enterprise security infrastructure, such as identity management, access control, and data governance systems, allowing organizations to reuse existing policies and security practices while adding controls designed for autonomous agent behavior.

The separation between policy management and runtime enforcement further enables flexible deployment across diverse AI agent environments. Since sidecars operate outside agent implementations, the same security layer can be applied across different models, frameworks, and application architectures without requiring modifications to individual agents. Sidecars can be deployed at agent interaction points or data boundaries between agents and protected resources, enabling organizations to enforce policies even when agents have heterogeneous internal designs or expose only black-box interfaces.

\section{Discussion \& Future Directions}

Our reference architecture is intended to motivate further exploration of the design space rather than serve as a final solution. It represents a step toward more secure AI agent systems while leaving several important challenges and limitations open for future research.

\textbf{Limitations.} Unlike existing agent-level defenses that rely on the agent itself to reason about and enforce privacy and security policies, our architecture delegates these decisions to an independent semantic engine running in a sidecar. This separation reduces the likelihood that policy enforcement is compromised by attacks on the agent. However, it does not eliminate the risk entirely. For example, an attacker may manipulate the context supplied by the agent, causing the semantic engine to authorize an otherwise unauthorized information flow or action. Second, our architecture is designed to enforce policies at the enterprise boundary. It does not provide visibility or control over computation performed by external services once data leaves the enterprise. Therefore, it should be complemented with trusted external services or additional safeguards for outbound data and requests.


\textbf{From egress to ingress.} The natural next step is to extend our egress-only mechanism to ingress, so that we filter not just what an agent sends but also what it receives. This cleanly separates the two directions. However, ingress is much harder. To inspect incoming traffic, the sidecar needs cooperation it cannot rely on: a malicious sender can simply avoid any channel that lets the sidecar look inside. And the information needed to enforce policy must now come from the sender, who may hide, drop, or fake it. Solving this likely means changing strategy—instead of inspecting traffic after it arrives, we would rely on senders to attach verifiable labels agreed upon ahead of time.

\textbf{Semantic policies and context.} A third set of questions concerns policy and context. How should semantic security policies be defined, and what context should they evaluate? Policies should clearly specify how information may be used while remaining precise, auditable, and adaptable to organization-specific requirements. Similarly, context representations should preserve the information and execution history needed for policy decisions without becoming too complex to enforce efficiently and consistently.


\textbf{Generalizability.} 
Deterministic enforcement works well when the set of agents is known in advance and workflow paths can be specified by predefined policies. In practice, however, AI agents often spawn sub-agents, invoke new tools, and establish previously unseen interactions at runtime. Supporting these dynamic behaviors requires more than static rules and remains an important area for future research. Dynamic enforcement mechanisms that complement predefined policies could help close this gap. 

Similarly, supporting a dynamic role-based system introduces a second challenge. At startup, each agent registers with the control plane by declaring its role and trust level, from which the control plane derives its capabilities according to role-based policies. However, roles, trust levels, and capabilities may evolve as agents execute tasks and interact with new tools. Continuously reevaluating these assignments improves security but incurs additional overhead. Balancing timely policy updates with efficient enforcement remains an open research challenge.

Finally, the policy router (Figure~\ref{fig:execution-flow}) raises an important research question: how can enforcement resources be allocated based on risk? A risk-aware router could direct low-risk flows through efficient deterministic checks while reserving more expensive context-aware analysis for higher-risk flows. Designing accurate and adaptive risk classification mechanisms to support this tradeoff remains an open direction for future work.



\bibliographystyle{plain}
\bibliography{refs}

@inproceedings{casado2006sane,
  title     = {{SANE}: A Protection Architecture for Enterprise Networks},
  author    = {Casado, Martin and Garfinkel, Tal and Akella, Aditya and Freedman, Michael J. and Boneh, Dan and McKeown, Nick and Shenker, Scott},
  booktitle = {15th USENIX Security Symposium (USENIX Security 06)},
  year      = {2006},
  address   = {Vancouver, B.C., Canada}
}

@inproceedings{casado2007ethane,
  title     = {Ethane: Taking Control of the Enterprise},
  author    = {Casado, Martin and Freedman, Michael J. and Pettit, Justin and Luo, Jianying and McKeown, Nick and Shenker, Scott},
  booktitle = {Proceedings of the ACM SIGCOMM 2007 Conference},
  pages     = {1--12},
  year      = {2007},
  doi       = {10.1145/1282380.1282382}
}

@article{saltzer1984e2e,
  title   = {End-to-End Arguments in System Design},
  author  = {Saltzer, Jerome H. and Reed, David P. and Clark, David D.},
  journal = {ACM Transactions on Computer Systems},
  volume  = {2},
  number  = {4},
  pages   = {277--288},
  year    = {1984},
  doi     = {10.1145/357401.357402}
}

@inproceedings{yaar2004siff,
  title     = {{SIFF}: A Stateless Internet Flow Filter to Mitigate {DDoS} Flooding Attacks},
  author    = {Yaar, Abraham and Perrig, Adrian and Song, Dawn},
  booktitle = {2004 IEEE Symposium on Security and Privacy},
  pages     = {130--143},
  year      = {2004},
  doi       = {10.1109/SECPRI.2004.1301320}
}

@inproceedings{yang2005tva,
  title     = {A {DoS}-Limiting Network Architecture},
  author    = {Yang, Xiaowei and Wetherall, David and Anderson, Thomas},
  booktitle = {Proceedings of the ACM SIGCOMM 2005 Conference},
  pages     = {241--252},
  year      = {2005},
  doi       = {10.1145/1080091.1080120}
}

@inproceedings{ballani2005offbydefault,
  title     = {Off by Default!},
  author    = {Ballani, Hitesh and Chawathe, Yatin and Ratnasamy, Sylvia and Roscoe, Timothy and Shenker, Scott},
  booktitle = {Proceedings of the 4th ACM Workshop on Hot Topics in Networks (HotNets-IV)},
  year      = {2005},
  address   = {College Park, MD}
}

@inproceedings{shao2024privacylens,
  title     = {{P}rivacy{L}ens: Evaluating Privacy Norm Awareness of Language Models in Action},
  author    = {Shao, Yijia and Li, Tianshi and Shi, Weiyan and Liu, Yanchen and Yang, Diyi},
  booktitle = {Advances in Neural Information Processing Systems},
  volume    = {37},
  series    = {NeurIPS Datasets and Benchmarks Track},
  year      = {2024},
  eprint    = {2409.00138},
  archivePrefix = {arXiv},
  primaryClass  = {cs.CL},
  url       = {https://arxiv.org/abs/2409.00138}
}

@inproceedings{zharmagambetov2025agentdam,
  title     = {{A}gent{DAM}: Privacy Leakage Evaluation for Autonomous Web Agents},
  author    = {Zharmagambetov, Arman and Guo, Chuan and Evtimov, Ivan and Pavlova, Maya and Salakhutdinov, Ruslan and Chaudhuri, Kamalika},
  booktitle = {Advances in Neural Information Processing Systems},
  volume    = {39},
  year      = {2025},
  eprint    = {2503.09780},
  archivePrefix = {arXiv},
  primaryClass  = {cs.AI},
  url       = {https://arxiv.org/abs/2503.09780}
}

@inproceedings{jia2024taskshield,
  title     = {The Task Shield: Enforcing Task Alignment to Defend Against Indirect Prompt Injection in {LLM} Agents},
  author    = {Jia, Feiran and Wu, Tong and Qin, Xin and Squicciarini, Anna},
  booktitle = {Proceedings of the 63rd Annual Meeting of the Association for Computational Linguistics (Volume 1: Long Papers)},
  pages     = {29680--29697},
  year      = {2025},
  month     = jul,
  address   = {Vienna, Austria},
  publisher = {Association for Computational Linguistics},
  doi       = {10.18653/v1/2025.acl-long.1435},
  eprint    = {2412.16682},
  archivePrefix = {arXiv},
  primaryClass  = {cs.CR}
}

@inproceedings{bagdasarian2024airgapagent,
  title     = {{A}ir{G}ap{A}gent: Protecting Privacy-Conscious Conversational Agents},
  author    = {Bagdasarian, Eugene and Yi, Ren and Ghalebikesabi, Sahra and Kairouz, Peter and Gruteser, Marco and Oh, Sewoong and Balle, Borja and Ramage, Daniel},
  booktitle = {Proceedings of the 2024 ACM SIGSAC Conference on Computer and Communications Security},
  series    = {CCS '24},
  pages     = {3868--3882},
  year      = {2024},
  address   = {New York, NY, USA},
  publisher = {Association for Computing Machinery},
  isbn      = {9798400706363},
  doi       = {10.1145/3658644.3690350},
  eprint    = {2405.05175},
  archivePrefix = {arXiv},
  primaryClass  = {cs.CR}
}

@article{elyagoubi2026agentleak,
  title   = {{A}gent{L}eak: A Full-Stack Benchmark for Privacy Leakage in Multi-Agent {LLM} Systems},
  author  = {El Yagoubi, Faouzi and Badu-Marfo, Godwin and Al Mallah, Ranwa},
  journal = {arXiv preprint arXiv:2602.11510},
  year    = {2026},
  month   = feb,
  eprint  = {2602.11510},
  archivePrefix = {arXiv},
  primaryClass  = {cs.AI},
  url     = {https://arxiv.org/abs/2602.11510},
  note    = {Preprint}
}

@inproceedings{wang2025privacyinaction,
  title     = {Privacy in Action: Towards Realistic Privacy Mitigation and Evaluation for {LLM}-Powered Agents},
  author    = {Wang, Shouju and Yu, Fenglin and Liu, Xirui and Qin, Xiaoting and Zhang, Jue and Lin, Qingwei and Zhang, Dongmei and Rajmohan, Saravan},
  booktitle = {Findings of the Association for Computational Linguistics: EMNLP 2025},
  pages     = {17055--17074},
  year      = {2025},
  month     = nov,
  address   = {Suzhou, China},
  publisher = {Association for Computational Linguistics},
  isbn      = {979-8-89176-335-7},
  doi       = {10.18653/v1/2025.findings-emnlp.925},
  eprint    = {2509.17488},
  archivePrefix = {arXiv},
  primaryClass  = {cs.CR}
}

@inproceedings{lan2025cirl,
  title     = {Contextual Integrity in {LLM}s via Reasoning and Reinforcement Learning},
  author    = {Lan, Guangchen and Inan, Huseyin A. and Abdelnabi, Sahar and Kulkarni, Janardhan and Wutschitz, Lukas and Shokri, Reza and Brinton, Christopher G. and Sim, Robert},
  booktitle = {Advances in Neural Information Processing Systems},
  volume    = {39},
  year      = {2025},
  eprint    = {2506.04245},
  archivePrefix = {arXiv},
  primaryClass  = {cs.AI},
  url       = {https://arxiv.org/abs/2506.04245}
}

@inproceedings{sherry2012aplomb,
  author    = {Sherry, Justine and Hasan, Shaddi and Scott, Colin and Krishnamurthy, Arvind and Ratnasamy, Sylvia and Sekar, Vyas},
  title     = {Making Middleboxes Someone Else's Problem: Network Processing as a Cloud Service},
  booktitle = {ACM SIGCOMM},
  year      = {2012}
}

@inproceedings{sherry2015blindbox,
  author    = {Sherry, Justine and Lan, Chang and Popa, Raluca Ada and Ratnasamy, Sylvia},
  title     = {{BlindBox}: Deep Packet Inspection over Encrypted Traffic},
  booktitle = {ACM SIGCOMM},
  year      = {2015}
}

@inproceedings{sherry2015ftmb,
  author    = {Sherry, Justine and Gao, Peter Xiang and Basu, Soumya and Panda, Aurojit and Krishnamurthy, Arvind and Maciocco, Christian and Manesh, Maziar and Martins, Jo{\~a}o and Ratnasamy, Sylvia and Rizzo, Luigi and Shenker, Scott},
  title     = {Rollback-Recovery for Middleboxes},
  booktitle = {ACM SIGCOMM},
  year      = {2015}
}

@inproceedings{myers1997dlm,
  author    = {Myers, Andrew C. and Liskov, Barbara},
  title     = {A Decentralized Model for Information Flow Control},
  booktitle = {ACM SOSP},
  year      = {1997}
}

@inproceedings{zeldovich2006histar,
  author    = {Zeldovich, Nickolai and Boyd-Wickizer, Silas and Kohler, Eddie and Mazi{\`e}res, David},
  title     = {Making Information Flow Explicit in {HiStar}},
  booktitle = {USENIX OSDI},
  year      = {2006}
}

@inproceedings{krohn2007flume,
  author    = {Krohn, Maxwell and Yip, Alexander and Brodsky, Micah and Cliffer, Natan and Kaashoek, M. Frans and Kohler, Eddie and Morris, Robert},
  title     = {Information Flow Control for Standard {OS} Abstractions},
  booktitle = {ACM SOSP},
  year      = {2007}
}

@article{nissenbaum2004ci,
  author  = {Nissenbaum, Helen},
  title   = {Privacy as Contextual Integrity},
  journal = {Washington Law Review},
  volume  = {79},
  number  = {1},
  pages   = {119--157},
  year    = {2004}
}

@inproceedings{mireshghallah2024confaide,
  author    = {Mireshghallah, Niloofar and Kim, Hyunwoo and Zhou, Xuhui and Tsvetkov, Yulia and Sap, Maarten and Shokri, Reza and Choi, Yejin},
  title     = {Can {LLMs} Keep a Secret? {T}esting Privacy Implications of Language Models via Contextual Integrity Theory},
  booktitle = {ICLR},
  year      = {2024}
}

@inproceedings{greshake2023injection,
  author    = {Greshake, Kai and Abdelnabi, Sahar and Mishra, Shailesh and Endres, Christoph and Holz, Thorsten and Fritz, Mario},
  title     = {Not What You've Signed Up For: Compromising Real-World {LLM}-Integrated Applications with Indirect Prompt Injection},
  booktitle = {ACM Workshop on Artificial Intelligence and Security (AISec)},
  year      = {2023}
}

@inproceedings{debenedetti2024agentdojo,
  author    = {Debenedetti, Edoardo and Zhang, Jie and Balunovi{\'c}, Mislav and Beurer-Kellner, Luca and Fischer, Marc and Tram{\`e}r, Florian},
  title     = {{AgentDojo}: A Dynamic Environment to Evaluate Prompt Injection Attacks and Defenses for {LLM} Agents},
  booktitle = {NeurIPS Datasets and Benchmarks},
  year      = {2024}
}

@inproceedings{yao2023react,
  author    = {Yao, Shunyu and Zhao, Jeffrey and Yu, Dian and Du, Nan and Shafran, Izhak and Narasimhan, Karthik and Cao, Yuan},
  title     = {{ReAct}: Synergizing Reasoning and Acting in Language Models},
  booktitle = {ICLR},
  year      = {2023}
}

@misc{mcp2024,
  author       = {Anthropic},
  title        = {Model Context Protocol},
  howpublished = {\url{https://modelcontextprotocol.io}},
  year         = {2024}
}

@misc{a2a2025,
  author       = {Google},
  title        = {Agent2Agent ({A2A}) Protocol},
  howpublished = {\url{https://a2a-protocol.org}},
  year         = {2025}
}

@inproceedings{rebedea2023nemo,
  title={{NeMo} Guardrails: A Toolkit for Controllable and Safe {LLM} Applications with Programmable Rails},
  author={Rebedea, Traian and Dinu, Razvan and Sreedhar, Makesh Narsimhan and Parisien, Christopher and Cohen, Jonathan},
  booktitle={Proceedings of the 2023 Conference on Empirical Methods in Natural Language Processing: System Demonstrations},
  pages={431--445},
  year={2023}
}

@article{inan2023llamaguard,
  title={Llama Guard: LLM-based Input-Output Safeguard for Human-AI Conversations},
  author={Inan, Hakan and Upasani, Kartikeya and Chi, Jianfeng and Rungta, Rashi and Iyer, Krithika and Mao, Yuning and Tontchev, Michael and Hu, Qing and Fuller, Brian and Testuggine, Davide and Khabsa, Madian},
  journal={arXiv preprint arXiv:2312.06674},
  year={2023}
}

@article{sandhu1996rbac,
  title   = {Role-Based Access Control Models},
  author  = {Sandhu, Ravi S. and Coyne, Edward J. and Feinstein, Hal L. and Youman, Charles E.},
  journal = {Computer},
  volume  = {29},
  number  = {2},
  pages   = {38--47},
  year    = {1996},
  publisher = {IEEE},
  doi     = {10.1109/2.485845}
}

@techreport{hardt2012oauth,
  title       = {The {OAuth} 2.0 Authorization Framework},
  author      = {Hardt, Dick},
  institution = {Internet Engineering Task Force (IETF)},
  number      = {RFC 6749},
  year        = {2012},
  month       = oct,
  url         = {https://www.rfc-editor.org/rfc/rfc6749}
}

@inproceedings{zhou2025rescriber,
  title     = {{R}escriber: Smaller-{LLM}-Powered User-Led Data Minimization for {LLM}-Based Chatbots},
  author    = {Zhou, Jijie and Xu, Eryue and Wu, Yaoyao and Li, Tianshi},
  booktitle = {Proceedings of the 2025 CHI Conference on Human Factors in Computing Systems},
  series    = {CHI '25},
  pages     = {1--28},
  year      = {2025},
  month     = apr,
  address   = {Yokohama, Japan},
  publisher = {Association for Computing Machinery},
  isbn      = {979-8-4007-1394-1},
  doi       = {10.1145/3706598.3713701},
  eprint    = {2410.11876},
  archivePrefix = {arXiv},
  primaryClass  = {cs.HC}
}

@misc{lakera2023guard,
  title        = {Lakera Guard},
  author       = {{Lakera AI}},
  year         = {2023},
  howpublished = {\url{https://www.lakera.ai/lakera-guard}},
  note         = {Accessed 2026-07-08}
}

@article{saltzer1975protection,
  title   = {The Protection of Information in Computer Systems},
  author  = {Saltzer, Jerome H. and Schroeder, Michael D.},
  journal = {Proceedings of the IEEE},
  volume  = {63},
  number  = {9},
  pages   = {1278--1308},
  year    = {1975},
  month   = sep,
  publisher = {IEEE},
  doi     = {10.1109/PROC.1975.9939}
}

@article{denning1987intrusion,
  title   = {An Intrusion-Detection Model},
  author  = {Denning, Dorothy E.},
  journal = {IEEE Transactions on Software Engineering},
  volume  = {SE-13},
  number  = {2},
  pages   = {222--232},
  year    = {1987},
  month   = feb,
  publisher = {IEEE},
  doi     = {10.1109/TSE.1987.232894}
}

@techreport{rose2020zerotrust,
  title       = {Zero Trust Architecture},
  author      = {Rose, Scott and Borchert, Oliver and Mitchell, Stu and Connelly, Sean},
  institution = {National Institute of Standards and Technology (NIST)},
  number      = {NIST Special Publication (SP) 800-207},
  year        = {2020},
  month       = aug,
  address     = {Gaithersburg, MD},
  doi         = {10.6028/NIST.SP.800-207}
}

@article{chen2024agentpoison,
  title={Agentpoison: Red-teaming llm agents via poisoning memory or knowledge bases},
  author={Chen, Zhaorun and Xiang, Zhen and Xiao, Chaowei and Song, Dawn and Li, Bo},
  journal={Advances in Neural Information Processing Systems},
  volume={37},
  pages={130185--130213},
  year={2024}
}

@article{dong2026memory,
  title={Memory injection attacks on LLM agents via query-only interaction},
  author={Dong, Shen and Xu, Shaochen and He, Pengfei and Li, Yige and Tang, Jiliang and Liu, Tianming and Liu, Hui and Xiang, Zhen},
  journal={Advances in Neural Information Processing Systems},
  volume={38},
  pages={46697--46731},
  year={2026}
}

@article{yazdinejad2026temporal,
  title={Temporal Dynamics of Memory Poisoning in Web3-Style LLM Agents},
  author={Yazdinejad, Abbas and Karimipour, Hadis},
  journal={IEEE Access},
  year={2026},
  publisher={IEEE}
}

@article{markov2023holistic,
  title={A Holistic Approach to Undesired Content Detection in the Real World},
  author={Markov, Todor and Zhang, Chong and Agarwal, Sandhini and Nekoul, Florentine Eloundou and Lee, Theodore and Adler, Steven and Jiang, Angela and Weng, Lilian},
  journal={Proceedings of the AAAI Conference on Artificial Intelligence},
  volume={37},
  number={12},
  pages={15009--15018},
  year={2023},
  doi={10.1609/aaai.v37i12.26752}
}

@article{deng2025ai,
  title={Ai agents under threat: A survey of key security challenges and future pathways},
  author={Deng, Zehang and Guo, Yongjian and Han, Changzhou and Ma, Wanlun and Xiong, Junwu and Wen, Sheng and Xiang, Yang},
  journal={ACM Computing Surveys},
  volume={57},
  number={7},
  pages={1--36},
  year={2025},
  publisher={ACM New York, NY}
}

@article{zhang2024towards,
  title={Towards action hijacking of large language model-based agent},
  author={Zhang, Yuyang and Chen, Kangjie and Gao, Jiaxin and Cui, Ronghao and Wang, Run and Wang, Lina and Zhang, Tianwei},
  journal={arXiv preprint arXiv:2412.10807},
  year={2024}
}

@article{wallace2024instruction,
  title={The Instruction Hierarchy: Training LLMs to Prioritize Privileged Instructions},
  author={Wallace, Eric and Xiao, Kai and Leike, Reimar and Weng, Lilian and Heidecke, Johannes and Beutel, Alex},
  journal={arXiv preprint arXiv:2404.13208},
  year={2024}
}

@article{jha2025breaking,
  title={Breaking and Fixing Defenses Against Control-Flow Hijacking in Multi-Agent Systems},
  author={Jha, Rishi and Triedman, Harold and Wagle, Justin and Shmatikov, Vitaly},
  journal={arXiv preprint arXiv:2510.17276},
  year={2025}
}

@inproceedings{zhan2024injecagent,
  title={Injecagent: Benchmarking indirect prompt injections in tool-integrated large language model agents},
  author={Zhan, Qiusi and Liang, Zhixiang and Ying, Zifan and Kang, Daniel},
  booktitle={Findings of the Association for Computational Linguistics: ACL 2024},
  pages={10471--10506},
  year={2024}
}

@article{evtimov2026wasp,
  title={Wasp: Benchmarking web agent security against prompt injection attacks},
  author={Evtimov, Ivan and Zharmagambetov, Arman and Grattafiori, Aaron and Guo, Chuan and Chaudhuri, Kamalika},
  journal={Advances in Neural Information Processing Systems},
  volume={38},
  year={2026}
}

@article{alizadeh2025simple,
  title={Simple prompt injection attacks can leak personal data observed by llm agents during task execution},
  author={Alizadeh, Meysam and Samei, Zeynab and Stetsenko, Daria and Gilardi, Fabrizio},
  journal={arXiv preprint arXiv:2506.01055},
  year={2025}
}

@article{naik2026omni,
  title={OMNI-LEAK: Orchestrator multi-agent network induced data leakage},
  author={Naik, Akshat and Culligan, Jay and Gal, Yarin and Torr, Philip and Aljundi, Rahaf and Paren, Alasdair and Bibi, Adel},
  journal={arXiv preprint arXiv:2602.13477},
  year={2026}
}

@article{louck2025improving,
  title={Improving Google A2A protocol: Protecting sensitive data and mitigating unintended harms in multi-agent systems},
  author={Louck, Yedidel and Stulman, Ariel and Dvir, Amit},
  journal={ACM Transactions on Software Engineering and Methodology},
  year={2025},
  publisher={ACM New York, NY}
}

@article{tan2026prompt,
  title={From Prompt Injection to Persistent Control: Defending Agentic Harness Against Trojan Backdoors},
  author={Tan, Jiejun and Dou, Zhicheng and Yang, Xinyu and Hu, Yuyang and Cheng, Yiruo and Li, Xiaoxi and Wen, Ji-Rong},
  journal={arXiv preprint arXiv:2605.31042},
  year={2026}
}

@article{zhu2025cve,
  title={CVE-bench: a benchmark for AI agents' ability to exploit real-world web application vulnerabilities},
  author={Zhu, Yuxuan and Kellermann, Antony and Bowman, Dylan and Li, Philip and Gupta, Akul and Danda, Adarsh and Fang, Richard and Jensen, Conner and Ihli, Eric and Benn, Jason and others},
  journal={arXiv preprint arXiv:2503.17332},
  year={2025}
}

@misc{csa2026overprivileged,
  author       = {{Cloud Security Alliance AI Safety Initiative}},
  title        = {Overprivileged by Design: {AI} Agents as Cloud Escalation Vectors},
  howpublished = {Lab Space, Cloud Security Alliance},
  year         = {2026},
  month        = apr,
  url          = {https://labs.cloudsecurityalliance.org/research/csa-research-note-ai-agent-cloud-privilege-escalation-202604/},
  urldate      = {2026-07-15},
  note         = {CSA Research Note}
}

@misc{thehackernews2026authbypass,
  author       = {{The Hacker News}},
  title        = {{AI} Agents Are Becoming Authorization Bypass Paths},
  howpublished = {The Hacker News},
  year         = {2026},
  month        = jan,
  url          = {https://thehackernews.com/2026/01/ai-agents-are-becoming-privilege.html},
  urldate      = {2026-07-15}
}

@misc{freestone2026aiagents,
  author       = {Freestone, Tim},
  title        = {{AI} Agents Are the Biggest Data Security Threat You're Not Governing},
  howpublished = {Kiteworks, Cybersecurity Risk Management Blog},
  year         = {2026},
  month        = feb,
  url          = {https://www.kiteworks.com/cybersecurity-risk-management/ai-agents-ungoverned-data-security-threat/},
  urldate      = {2026-07-15},
  note         = {Updated February 26, 2026}
}

@article{patlan2025real,
  title={Real ai agents with fake memories: Fatal context manipulation attacks on web3 agents},
  author={Patlan, Atharv Singh and Sheng, Peiyao and Hebbar, S Ashwin and Mittal, Prateek and Viswanath, Pramod},
  journal={arXiv preprint arXiv:2503.16248},
  year={2025}
}

@misc{oswal2026nondeterministic,
  author       = {Oswal, Anand},
  title        = {Why Traditional Security Fails in the Age of Non-Deterministic {AI}},
  howpublished = {Perspectives, Palo Alto Networks},
  year         = {2026},
  month        = feb,
  url          = {https://www.paloaltonetworks.com/perspectives/why-traditional-security-fails-in-the-age-of-non-deterministic-ai/},
  urldate      = {2026-07-15}
}

@article{jia2025critical,
  title={A critical evaluation of defenses against prompt injection attacks},
  author={Jia, Yuqi and Shao, Zedian and Liu, Yupei and Jia, Jinyuan and Song, Dawn and Gong, Neil Zhenqiang},
  journal={arXiv preprint arXiv:2505.18333},
  year={2025}
}

@inproceedings{chen2025struq,
  title={$\{$StruQ$\}$: Defending against prompt injection with structured queries},
  author={Chen, Sizhe and Piet, Julien and Sitawarin, Chawin and Wagner, David},
  booktitle={34th USENIX Security Symposium (USENIX Security 25)},
  pages={2383--2400},
  year={2025}
}

@inproceedings{chen2025secalign,
  title={Secalign: Defending against prompt injection with preference optimization},
  author={Chen, Sizhe and Zharmagambetov, Arman and Mahloujifar, Saeed and Chaudhuri, Kamalika and Wagner, David and Guo, Chuan},
  booktitle={Proceedings of the 2025 ACM SIGSAC Conference on Computer and Communications Security},
  pages={2833--2847},
  year={2025}
}

@misc{mccauley2025samemodel,
  author       = {McCauley, Conor and Schulz, Kasimir},
  title        = {Same Model, Different Hat},
  howpublished = {HiddenLayer Research},
  year         = {2025},
  month        = oct,
  url          = {https://www.hiddenlayer.com/research/same-model-different-hat},
  urldate      = {2026-07-15}
}

@inproceedings{wang2026sok,
  title={Sok: Evaluating jailbreak guardrails for large language models},
  author={Wang, Xunguang and Ji, Zhenlan and Wang, Wenxuan and Li, Zongjie and Wu, Daoyuan and Wang, Shuai},
  booktitle={2026 IEEE Symposium on Security and Privacy (SP)},
  pages={39--58},
  year={2026},
  organization={IEEE}
}

@misc{hackett2025bypassing,
  author        = {Hackett, William and Birch, Lewis and Trawicki, Stefan and Suri, Neeraj and Garraghan, Peter},
  title         = {Bypassing {LLM} Guardrails: An Empirical Analysis of Evasion Attacks against Prompt Injection and Jailbreak Detection Systems},
  year          = {2025},
  eprint        = {2504.11168},
  archivePrefix = {arXiv},
  primaryClass  = {cs.CR},
  url           = {https://arxiv.org/abs/2504.11168}
}

@article{lu2026clawless,
  title={ClawLess: A Security Model of AI Agents},
  author={Lu, Hongyi and Liu, Nian and Wang, Shuai and Zhang, Fengwei},
  journal={arXiv preprint arXiv:2604.06284},
  year={2026}
}

@article{narajala2025securing,
  title={Securing agentic ai: A comprehensive threat model and mitigation framework for generative ai agents},
  author={Narajala, Vineeth Sai and Narayan, Om},
  journal={arXiv preprint arXiv:2504.19956},
  year={2025}
}

@inproceedings{10.1145/3786582.3786839,
author = {Doshi, Aarya and Hong, Yining and Xu, Congying and Kang, Eunsuk and Kapravelos, Alexandros and K\"{a}stner, Christian},
title = {Towards Verifiably Safe Tool Use for LLM Agents},
year = {2026},
isbn = {9798400724251},
publisher = {Association for Computing Machinery},
address = {New York, NY, USA},
url = {https://doi.org/10.1145/3786582.3786839},
doi = {10.1145/3786582.3786839},
booktitle = {Proceedings of the IEEE/ACM 48th International Conference on Software Engineering},
pages = {201–205},
numpages = {5},
location = {
},
series = {ICSE-NIER '26}
}

@article{luo2026agentauditor,
  title={Agentauditor: Human-level safety and security evaluation for llm agents},
  author={Luo, Hanjun and Dai, Shenyu and Ni, Chiming and Li, Xinfeng and Zhang, Guibin and Wang, Kun and Liu, Tongliang and Salam, Hanan},
  journal={Advances in Neural Information Processing Systems},
  volume={38},
  pages={43241--43298},
  year={2026}
}

@article{qu2026overeager,
  title={Overeager coding agents: Measuring out-of-scope actions on benign tasks},
  author={Qu, Yubin and Zhang, Ying and Zhang, Yanjun and Deng, Gelei and Li, Yuekang and Zhang, Leo Yu and Liu, Yi},
  journal={arXiv preprint arXiv:2605.18583},
  year={2026}
}

@article{jones2026benign,
  title={When benign inputs lead to severe harms: Eliciting unsafe unintended behaviors of computer-use agents},
  author={Jones, Jaylen and Zhang, Zhehao and Ning, Yuting and Fosler-Lussier, Eric and St-Charles, Pierre-Luc and Bengio, Yoshua and Song, Dawn and Su, Yu and Sun, Huan},
  journal={arXiv preprint arXiv:2602.08235},
  year={2026}
}

@article{wu2024isolategpt,
  title={Isolategpt: An execution isolation architecture for llm-based agentic systems},
  author={Wu, Yuhao and Roesner, Franziska and Kohno, Tadayoshi and Zhang, Ning and Iqbal, Umar},
  journal={arXiv preprint arXiv:2403.04960},
  year={2024}
}

@article{debenedetti2025defeating,
  title={Defeating prompt injections by design},
  author={Debenedetti, Edoardo and Shumailov, Ilia and Fan, Tianqi and Hayes, Jamie and Carlini, Nicholas and Fabian, Daniel and Kern, Christoph and Shi, Chongyang and Terzis, Andreas and Tram{\`e}r, Florian},
  journal={arXiv preprint arXiv:2503.18813},
  year={2025}
}

@article{meng2025cellmate,
  title={cellmate: Sandboxing browser ai agents},
  author={Meng, Luoxi and Feng, Henry and Shumailov, Ilia and Fernandes, Earlence},
  journal={arXiv preprint arXiv:2512.12594},
  year={2025}
}

@article{piao2025agentbay,
  title={AgentBay: A Hybrid Interaction Sandbox for Seamless Human-AI Intervention in Agentic Systems},
  author={Piao, Yun and Min, Hongbo and Su, Hang and Zhang, Leilei and Wang, Lei and Yin, Yue and Wu, Xiao and Xu, Zhejing and Qu, Liwei and Li, Hang and others},
  journal={arXiv preprint arXiv:2512.04367},
  year={2025}
}

@article{shi2025progent,
  title={Progent: Programmable privilege control for llm agents},
  author={Shi, Tianneng and He, Jingxuan and Wang, Zhun and Wu, Linyu and Li, Hongwei and Guo, Wenbo and Song, Dawn},
  journal={arXiv e-prints},
  pages={arXiv--2504},
  year={2025}
}

@article{costa2025securing,
  title={Securing ai agents with information-flow control},
  author={Costa, Manuel and K{\"o}pf, Boris and Kolluri, Aashish and Paverd, Andrew and Russinovich, Mark and Salem, Ahmed and Tople, Shruti and Wutschitz, Lukas and Zanella-B{\'e}guelin, Santiago},
  journal={arXiv preprint arXiv:2505.23643},
  year={2025}
}

@article{zhong2025rtbas,
  title={Rtbas: Defending llm agents against prompt injection and privacy leakage},
  author={Zhong, Peter Yong and Chen, Siyuan and Wang, Ruiqi and McCall, McKenna and Titzer, Ben L and Miller, Heather and Gibbons, Phillip B},
  journal={arXiv preprint arXiv:2502.08966},
  year={2025}
}

@article{miculicich2025veriguard,
  title={Veriguard: Enhancing llm agent safety via verified code generation},
  author={Miculicich, Lesly and Parmar, Mihir and Palangi, Hamid and Dvijotham, Krishnamurthy Dj and Montanari, Mirko and Pfister, Tomas and Le, Long T},
  journal={arXiv preprint arXiv:2510.05156},
  year={2025}
}

@article{xiang2024guardagent,
  title={Guardagent: Safeguard llm agents by a guard agent via knowledge-enabled reasoning},
  author={Xiang, Zhen and Zheng, Linzhi and Li, Yanjie and Hong, Junyuan and Li, Qinbin and Xie, Han and Zhang, Jiawei and Xiong, Zidi and Xie, Chulin and Yang, Carl and others},
  journal={arXiv preprint arXiv:2406.09187},
  year={2024}
}
\end{document}